\documentclass[longbibliography,twocolumn,aps,prb,showkeys,superscriptaddress,letterpaper]{revtex4-2}
\usepackage[utf8]{inputenc}
\usepackage{amsmath}
\usepackage{amssymb}
\usepackage{graphicx}
\graphicspath{{plots/}}
\usepackage{color}

\usepackage[unicode=true,
 bookmarks=true,bookmarksnumbered=false,bookmarksopen=false,
 breaklinks=false,pdfborder={0 0 0},pdfborderstyle={},backref=false,colorlinks=true,]
 {hyperref}
\hypersetup{linkcolor=blue,citecolor=blue,urlcolor=blue}

\begin{document}

\title{Probing Boundary Spins in the Su-Schrieffer-Heeger-Hubbard model}

\author{Armando A. Aligia}\email{aaligia@gmail.com}

\affiliation{Instituto de Nanociencia y Nanotecnolog\'{\i}a CNEA-CONICET, GAIDI,
Centro At\'{o}mico Bariloche and Instituto Balseiro, 8400 Bariloche, Argentina}

\author{Alejandro M. Lobos}\email{alejandro.martin.lobos@gmail.com}

\affiliation{Instituto Interdisciplinario de Ciencias B\'{a}sicas - Consejo Nacional de Investigaciones Cient\'{i}ficas y T\'{e}cnicas }
\affiliation{Facultad de Ciencias Exactas y Naturales - Universidad Nacional de Cuyo}

\author{Lucila Peralta Gavensky}

\affiliation{Center for Nonlinear Phenomena and Complex Systems, Universit\'e Libre de Bruxelles, CP 231, Campus Plaine, B-1050 Brussels, Belgium}
\affiliation{International Solvay Institutes, 1050 Brussels, Belgium}

\author{Claudio J. Gazza}

\affiliation{Instituto de F\'\i sica Rosario, CONICET, and Facultad de Ciencias Exactas, Ingenier\'\i a y Agrimensura,
\\
Universidad Nacional de Rosario, 2000 Rosario Argentina}

\begin{abstract}
Studying boundary excitations provides a powerful approach to probe correlations in topological phases. We propose that localized spins
near the ends of
a Su–Schrieffer–Heeger–Hubbard chain embedded in an insulating environment can be detected experimentally using scanning tunneling microscopy (STM) combined with electron spin resonance (ESR). When the STM tip is in the contact regime, the tip–end-spin coupling realizes an effective Anderson impurity problem, giving rise to a Kondo peak at low bias. Spatially resolving the Kondo resonance width as the STM tip approaches the chain ends provides an indirect yet clear signature of these localized spins. To support this proposal, we use density-matrix renormalization group (DMRG) to calculate the spin gap and spin projection of end states for chains of various lengths and interaction strengths $U$ at half-filling. In the non-interacting limit ($U=0$), we derive simple analytical expressions that reproduce the numerical results for sufficiently long chains. We also discuss how the correlated phase of the isolated chain is characterized by boundary zeros in its single-particle Green's function, and briefly comment on their localization properties in relation to the boundary spins.

\end{abstract}

\maketitle

\section{Introduction}

Non-interacting topological insulators are at present well understood.
They can be characterized by a non-vanishing topological invariant \cite{Ando13}.
In particular for the Su-Schrieffer-Heeger chain (SSHC) \cite{Su79},
the Zak Berry phase \cite{Zak89} acts as a $\mathbb{Z}_2$ number which
distinguishes the topological and non-topological phases \cite{Asb16}.
For an open non-interacting SSHC, the topological phase has one eigenstate
of zero energy for each spin, localized at each end of the chain, which imply poles
at zero energy in the Green’s function.

In the interacting case, the states at zero energy disappear, however the Berry phase
or similar topological invariants \cite{Aligia23} continue to separate
topological and non-topological phases. In particular, an invariant
based on the Green’s function at zero energy has been proposed \cite{Gura11,Man12}.
Several numerical studies have shown that the boundary poles of the non-interacting Green’s function
are transformed into zeros when correlations are included in the topological phase~\cite{Man12,Yoshida2014,Wagner23}. It has been argued that
these zeros reflect the presence of in-gap spin excitations (spinons)
at zero energy \cite{Yoshida2014,Wagner23}. The appearance of zeros in the Green’s functions is responsible for a topological transition in a
two-channel spin-1 Kondo model with easy-plane anisotropy which explains several relevant experiments for magnetic molecules on metallic substrates
\cite{Zitko21,Blesio23,BlesioB,Blesio24}. The importance of these zeros has also
been stressed in different contexts \cite{Blason2023,Luci23,WagnerB,Fabrizio23,Pangburn25,Flores25}
and particularly in relation with the spinons \cite{Wagner23,WagnerB,Fabrizio23}.
In the case of a SSHC with open boundary conditions in the topological phase, it
has been established that localized spins remain at the ends in the
presence of interactions~\cite{Man12},
but the relation between the localization length of these spin modes and that of the boundary zeros of the single-particle Green's function has not been addressed.
Moreover, the evolution of these localized states (in particular, their magnitude and spatial extent) from the non-interacting to the strongly-interacting limit as a function of the interaction parameter has not, to our knowledge, been thoroughly investigated.  Exploring these aspects would be of considerable interest.

From the experimental perspective, the detection of localized spins at the ends of interacting spin chains has attracted a lot of attention lately. In recent scanning tunneling microscopy (STM) experiments, the Kondo resonance at zero bias has been used to identify localized spins at the ends of topological chains deposited on metallic surfaces \cite{Mishra2021, Zhao24,Edens25}. In these works, the intensity of the atomically-resolved Kondo resonance has been  used to map the spin profile of the topological end states. However, for potential applications where a metallic environment is not desirable (e.g., spin qubits requiring long quantum coherence times $T_2$) the Kondo resonance becomes unobservable and other  detection tools are necessary. The recent advent of electron spin resonance combined with STM (ESR-STM) has enabled to extract locally-resolved dynamic information of such isolated spin systems, and has been successfully applied in spin-1/2 Ti clusters deposited on an insulating MgO decoupling layer \cite{Wang24}. However, while this technique enables the detection of localized end states in topological spin chains in non-metallic environments, it has been pointed out that its spectral resolution would be severely restricted in the case of long chains ($L > 30$ atoms) \cite{Mai24}. Therefore, complementary methods would be highly desirable.

\begin{figure}[t]
\includegraphics[width=0.95\columnwidth]{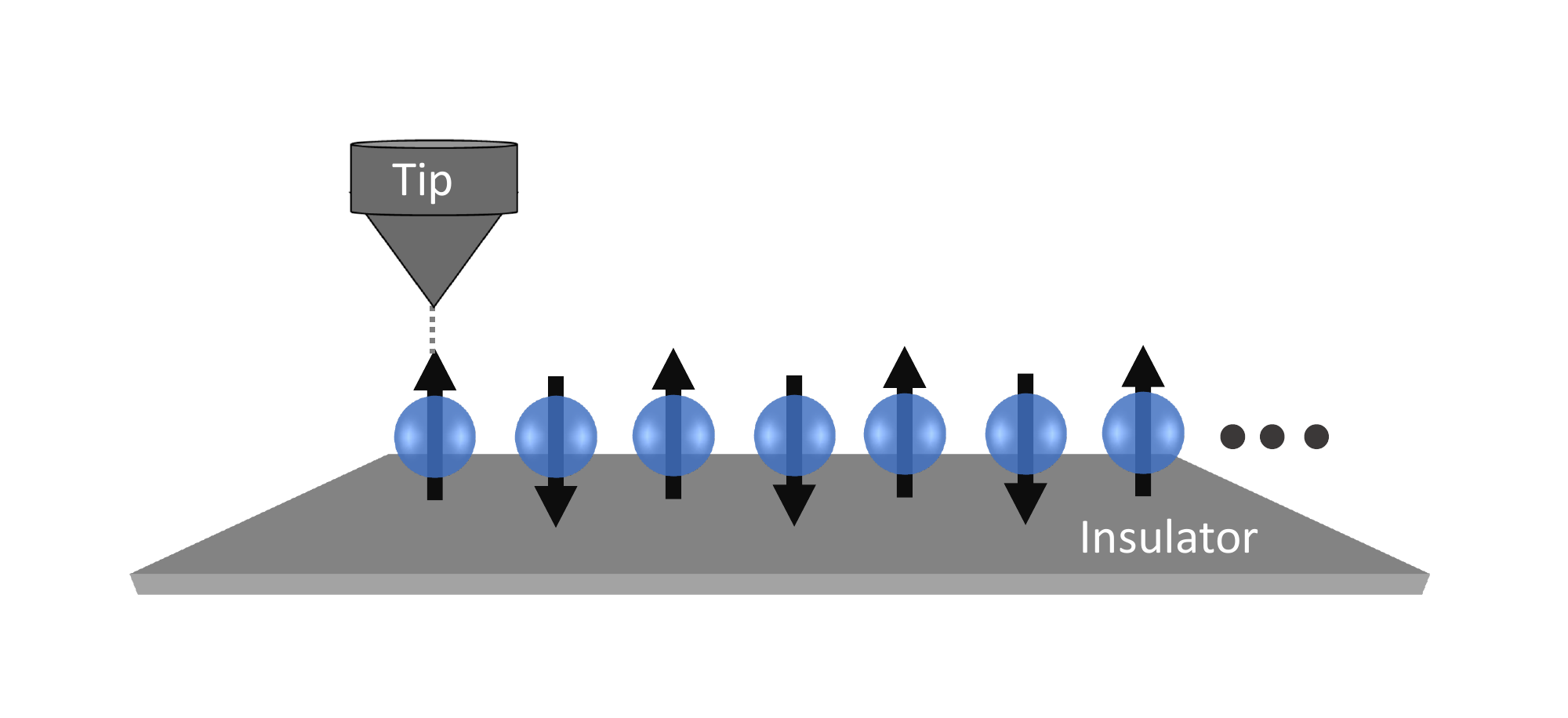}
\caption{(Color online) Scheme of an interacting spin chain on top of an insulating surface. In the contact regime, the conduction electrons in the STM tip can directly hybridize with the underlying localized spin states, effectively screening them.}
\label{system}
\end{figure}

In this article we
theoretically investigate how STM in the contact regime, possibly combined with ESR,
can provide critical insights into the localized spin degrees of freedom of this correlated insulator (see Fig. \ref{system}), thereby offering complementary information to standard STM-ESR spectroscopy. In fact it is known that magnetic molecules
on substrates can enter the contact regime when the STM tip is sufficiently
near to the molecule, favoring a direct  hybridization between the conduction electrons of the tip and
the localized electrons of the molecule. Examples are experiments
with nickelocene molecules on metallic surfaces
\cite{Ormaza17,Mohr20,Kogler24} and
the corresponding theory \cite{Blesio23}, and a Co atom on
a Cu(100) surface \cite{Choi16}.

Therefore, as explained in Section \ref{eff},
when the STM tip approaches an end of a SSHC in its topological phase,
the problem can be mapped into an effective impurity Anderson model, and
from the width (instead of the intensity) of the measured Kondo resonance, the magnitude
and degree of localization of the spin can be inferred. 
To that end, we propose an effective Hamiltonian for the interaction of the low-energy states of an interacting finite topological SSHC with an STM tip.
Using the density-matrix renormalization group (DMRG), we compute the spin gap and the expectation value of the spin projection
of several chains for different values of $U$, which serve as an input for the effective tip-system Hamiltonian.
Additionally, for a short SSHC, the spins at the ends interact leading
to a finite energy splitting, which can be detected by STM-ESR
\cite{Yang18,Willke18,Zhang22,Mai24}. For long chains naturally
an applied magnetic field also leads to a measurable splitting.
Recently both a Kondo resonance and ESR have been observed for
an organic molecule coupled to a metallic surface \cite{Chen25},
and ESR was used to study spin fractionalization in a Haldane
system \cite{Castillo24}.

The paper is organized as follows. Section \ref{model}
describes the SSHC model. Section \ref{zm} presents a simple analytical solution for the end states
of a non-interacting topological, open SSHC, starting from the exact zero-energy eigenstates of the infinite chain. Section \ref{eff} introduces the effective Hamiltonian relevant for the
ESR-STM experiments. Section \ref{res} contains the numerical DMRG results. Section \ref{green} discusses the emergence of zero eigenvalues of the zero-frequency single-particle Green's function and their localization properties. Finally, Section \ref{sum} provides a summary and discussion.

\section{Theoretical Model}

\label{model}

The SSHC was originally proposed to describe the physics of polyacetylene, where alternating strong and weak bonds between carbon atoms lead to a dimerized
lattice \cite{Su79}.
More recently, the interest in the system raised due to its topological
properties \cite{Asb16}. Two distinct topological states are separated
by a $\mathbb{Z}_2$ number \cite{Asb16,Aligia23} that changes if the strong
bonds are the even or the odd ones. The topological phase corresponds
to the case when the first bond of the chain is weak, and in this case
(as we show in detail in the next Section) the model has eigenstates of zero energy
localized at the ends of the chain.

In the interacting SSHC, the zero-energy states disappear, but the topological
properties of the model remain \cite{Aligia23,Man12} and spin excitations
localized at the ends persist in the topological phase.
In particular, in the strong-coupling limit, the model is equivalent
to a spin-1/2 alternating XXZ model \cite{Tzeng16,Marquez24},
and spin-1/2 edge excitations have been
detected in an experimental realization of this model using
graphene nanoribbons \cite{Zhao24}.

The Hamiltonian of the electron-hole symmetric interacting open SSHC, representing an atomic chain deposited on top of an insulating surface (see Fig. \ref{system}), can be
written in compact form as

\begin{eqnarray}
\hat{H} &=&\sum\limits_{j=1}^{L-1}-\left[ t_{0}+\delta \;(-1)^{j}\right]
\sum\limits_{\sigma ={\uparrow ,\downarrow }}\left( \hat{c}_{j+1\sigma }^{\dagger
}\hat{c}_{j\sigma }+\text{H.c.}\right)   \notag \\
&&+U\sum\limits_{j=1}^{L}\left( \hat{n}_{j\uparrow }\hat{n}_{j\downarrow }
-\frac{1}{2}\sum\limits_{\sigma ={\uparrow ,\downarrow }}\hat{n}_{j\sigma }\right),
\label{ham}
\end{eqnarray}
where $\hat{c}_{j\sigma }^{\dagger }$ creates an electron with spin $\sigma $ at
site $j$ and $\hat{n}_{j\sigma }=\hat{c}_{j\sigma }^{\dagger }\hat{c}_{j\sigma }$ is the electron number operator.
Unless otherwise stated, we assume that the number of sites
$L$ is even and the
system consists of $L/2$ unit cells. In the following we denote by
$t_{1}$ $(t_{2})$ the hopping of the odd (even) bonds:
\begin{equation}
t_{1}=t_{0}-\delta ,\text{ }t_{2}=t_{0}+\delta ,  \label{t12}
\end{equation}
which, without loss of generality, can be assumed positive (i.e., changing if necessary the phases of the
$c_{j\sigma }$ via a gauge transformation). The topological phase corresponds to the case $t_{1}<t_{2}$
\cite{Asb16,Man12,Aligia23}.

\section{States near zero energy for $U=0$}
\label{zm}

In the non-interacting case, the eigenstates of vanishing energy can be
easily found with the method of Alase \textit{et al.} \cite{Alase16,Alase17,Aligia18}. This
result sheds light of what is expected in the general case. The method is
particularly suitable to provide analytical results
when the energy is known (for example zero energy)
because self-consistent equations can be avoided \cite{Aligia18}.
In the topological phase of the non-interacting open SSCH for
$L\longrightarrow \infty $,  localized states with zero energy are
present at each end. Since it is easy to verify that these states are zero-energy
states, we just write the result we obtained  for the states localized at
the left ($|L\sigma \rangle $) and the right ($|R\sigma \rangle $) of the
infinite chain with spin $\sigma $. We can define these states as
\begin{eqnarray}
|L\sigma \rangle  &=&\hat{l}_\sigma^\dagger|0\rangle , \label{state_l}\\  
|R\sigma \rangle  &=&\hat{r}_\sigma^\dagger|0\rangle ,  \label{state_r}
\end{eqnarray}
where the second-quantized  operators $\hat{l}_\sigma^\dagger$ and $\hat{r}_\sigma^\dagger$ are defined as
\begin{eqnarray}
\hat{l}_\sigma^\dagger &=&\sqrt{\frac{1-z^{2}}{1-z^{L}}}
\sum\limits_{j=1}^{L/2}z^{j-1}\hat{c}_{2j-1\sigma }^{\dagger } ,  \label{op_lsigma}\\
\hat{r}_\sigma^\dagger &=&\sqrt{\frac{1-z^{2}}{1-z^{L}}}
\sum\limits_{j=1}^{L/2}z^{L/2-j}\hat{c}_{2j\sigma }^{\dagger },   \label{op_rsigma}\\
z &=&-t_{1}/t_{2}, 
\end{eqnarray}
where for later use we have generalized the definitions of the normalized states to chains of finite-length $L$.
Note that for $|L\sigma \rangle $  only odd sites enter in the sum with
decreasing amplitude $z$ in such a way that $H|L\sigma \rangle =0$.
Similarly only even sites enter $|R\sigma \rangle $. In fact the inversion
symmetry $I$ centered at the midpoint between sites $L/2$ and $L/2+1$, which
interchanges sites $j$ and $L-j+1$ ($\hat{c}_{j\sigma }^{\dagger }\leftrightarrow
\hat{c}_{L-j+1\sigma }^{\dagger }$) is a symmetry of the Hamiltonian ($[H,I]{}=0$)
and $|R\sigma \rangle =I|L\sigma \rangle $.

For a finite chain, although the method of Alase \textit{et al.} can be applied \cite{Aligia18},
we follow a simpler approach and start from Eqs. (\ref{state_l}) and (\ref{state_r}) as approximate
eigenstates. After some algebra we obtain

\begin{eqnarray}
H|L\sigma \rangle  &=&-t_{LR}|R\sigma \rangle +t_{c}|A\sigma \rangle ,
\notag \\
H|R\sigma \rangle  &=&-t_{LR}|L\sigma \rangle +t_{c}|B\sigma \rangle ,
\notag \\
t_{LR} &=&\frac{1-z^{2}}{1-z^{L}}z^{L/2-1}t_{1},  \label{2x2}
\end{eqnarray}
where all states are orthonormal, and states $|A\sigma \rangle$ and $|B\sigma \rangle$ are states in the continuum with energy above the gap, given by
\begin{eqnarray}
t_{c}|A\sigma \rangle  &=&\sqrt{\frac{1-z^{2}}{(1-z^{L})^{3}}}
z^{L/2-1}t_{1}[(1-z^{2})\sum\limits_{j=1}^{L/2-1}z^{L/2-j}c_{2j\sigma
}^{\dagger }  \notag \\
&&+(z^{L}-z^{2})c_{L\sigma }^{\dagger }]{}|0\rangle ,  \notag \\
|B\sigma \rangle  &=&I|A\sigma \rangle .  \label{ab}
\end{eqnarray}

As shown below, already for a chain of
moderate length such that $z^{L/2}\ll 1$, the terms proportional to $t_{c}$
[given by Eqs. (\ref{ab})] can be neglected and the eigenstates and
energies are

\begin{eqnarray}
|S\sigma\rangle &=&(|L\sigma \rangle +|R\sigma \rangle )/\sqrt{2}, {\rm } E_S=-t_{LR}   \notag \\
|D\sigma \rangle &=&(|L\sigma \rangle -|R\sigma \rangle )/\sqrt{2}, {\rm } E_D=t_{LR} \label{eigen}
\end{eqnarray}
The many-body ground state is a singlet with the
states $|S\sigma \rangle $ occupied for both spins. The lowest triplet state
has one electron occupying a  $|S\sigma \rangle $ and the other a  $|D\sigma
\rangle $ state. In particular for total spin projection $S_{z}=1$ the state
is simply $l_\uparrow^\dagger r_\uparrow^\dagger|0\rangle $.  Therefore the spin gap is

\begin{equation}
\Delta _{s}=2t_{LR}=2\frac{1-z^{2}}{1-z^{L}}z^{L/2-1}t_{1}.  \label{sg}
\end{equation}
This energy can be measured by the ESR-STM technique \cite{Wang24, Castillo24}.

Regarding the terms proportional to $t_{c}$ in Eq. (\ref{2x2}), note that from the norm of $t_{c}|A\sigma \rangle $, and neglecting $z^{L}$ compared
to 1, one has

\begin{equation}
t_{c}^{2}\simeq z^{L}(1-z^{2})t_{1}^{2}. \label{tc}
\end{equation}

The eigenstates
near zero energy are separated from the rest of the spectrum by an energy
$E=t_{2}-t_{1}$ (half the gap of the periodic non-interacting SSHC
\cite{Asb16}). Therefore, from second-order perturbation theory, the correction
of the energies of the low-energy eigenstates due to the states $|A\sigma
\rangle $ and $|B\sigma \rangle $ is of order
$t_{c}^{2}/E=z^{L+2}(t_{1}+t_{2})$
and the relative error in the spin gap is
of order
\begin{equation}
t_{c}^{2}/(E\Delta _{s})\simeq z^{L/2+1}t_{1}/(t_{2}-t_{1})\simeq z^{L/2+2}/(1+z).
\label{errorgap}
\end{equation}
Since we are assuming $z^L \ll 1$, we conclude that the terms proportional to $t_c$ can be safely neglected.

\section{Effective Hamiltonian for the ESR-STM}
\label{eff}

As schematically shown in Fig. \ref{system}, we now assume that the STM tip approaches one end of an open long SSHC, and we assume the tip is in the contact
regime \cite{Mohr20,Blesio23,Choi16}.
Under such conditions, the
localized end states can jump to the tip giving rise to a resonant level
model. Choosing for simplicity the left end, the Hamiltonian of the tip coupled to the spin-chain takes the form
\begin{equation}
\hat{H}_{1}(j)=\sum\limits_{k\sigma }[\varepsilon_{k\sigma }\hat{t}_{k\sigma }^{\dagger
}\hat{t}_{k\sigma }+V_{j}(\hat{l}_{\sigma }^{\dagger }\hat{t}_{k\sigma }+\text{H.c.})]{}\,,
\label{h1}
\end{equation}
where the first term describes a continuum of states of the STM tip giving rise to a featureless local density of states, and the
second term corresponds to the hybridization with the left end-state at site $j$. The value of $V_{j}$ depends on the position of the tip along the chain which, due to its atomic resolution, is proportional to the coefficient of $\hat{c}_{j\sigma }^{\dagger }$
in Eq. (\ref{op_lsigma}). For example, for a long non-interacting chain it is $\sqrt{1-z^{2}}V$ for $j=1$, 0 for $j=2$, $z^{2}\sqrt{1-z^{2}}V$ for $j=3$, etc., with $V$ a constant that depends on the distance between the STM tip and the sample.
The  tip-chain coupling can be modeled by
\begin{equation}
V_{j}=\sqrt{2|\langle \hat{S}_{j}^{z}\rangle |}V,  \label{vn}
\end{equation}
where $\hat{S}_{j}^{z}$ is the spin projection at site $j$, and where the expectation
value is taken in the lowest energy many-body state with total spin projection $S_{z}=1$, i.e.,  the state $\hat{l}_\uparrow^\dagger \hat{r}^\dagger_\uparrow |0\rangle$.
While Eq. (\ref{vn}) is analytically exact only in the non-interacting case and merely an educated guess for the interacting one, the numerical results presented in Section \ref{res} show (see Fig. \ref{fig:V2_vs_2Sz}) that it remains a remarkably accurate approximation in the latter.

The above model corresponds to an effective Anderson model, in which a localized
state with spin 1/2 (in our case the spin degenerate ground state of a
semi-infinite SSHC in its topological phase) is hybridized with
a conduction band (the metallic states of the STM tip).
The spectral density of the localized states in the non-interacting case is well known and consists of a peak centered at the
Fermi level, with the special feature of having a spatially-dependent width
$\Gamma_j =\rho V_{j}^{2}$, where $\rho $ is the spectral
density per spin of the tip, assumed independent of energy.
In contrast, the density of states of the conduction electrons has a dip
of the same magnitude $\Gamma_j$.

In the interacting case $U\neq 0$, the localized spin remains at the end.
The situation is particularly simple for $t_1=0$. In this case,
$\hat{c}_{1\sigma }^{\dagger } |0\rangle $ are eigenstates of the isolated
first site of the chain, with energy $-U/2$. Including the hybridization with the
tip, because of the electron-hole symmetry of our SSHC [Eq. \ref{ham}],
the model takes the form of the symmetric impurity-Anderson model (SIAM).

In the general interacting case of an infinite chain, there exist two degenerate states with total spin projection $S_{z}= \pm 1/2$, localized at the left end of the chain but extending beyond the first site. These states can be mapped onto the corresponding states of the SIAM (or Kondo model for large $U$),
and these two states can be flipped via second-order hopping to the STM tip.
The spin-flip terms are the primary source of the renormalization group
flow to strong coupling in the Kondo model \cite{Anderson70}.
Therefore we expect that the following SIAM is valid in the general case

\begin{equation}
\hat{H}_{2}(j)=\hat{H}_{1}(j)+U\left( \hat{n}_{L\uparrow }\hat{n}_{L\downarrow }
-\frac{1}{2}\sum\limits_{\sigma ={\uparrow ,\downarrow }}\hat{n}_{L\sigma }\right) ,
\label{ha}
\end{equation}
where  $\hat{n}_{L\sigma }=\hat{l}_{\sigma }^{\dagger }\hat{l}_{\sigma }$ for either $U=0$ or $t_1 \rightarrow 0$,
and the effective occupancy of the corresponding end state
in the general case.

The localized spectral density of the model Eq.~\eqref{ha} is well known. In the Kondo regime at low
energies it consists of a peak whose half width is of the order of the Kondo
temperature $T_{K}$. Using the Haldane formula \cite{Haldane78}
\begin{equation}
T_{K}(j)\simeq \sqrt{\frac{U\Gamma_j}{2} }\exp \left( \frac{-8\pi U}{\Gamma_j }\right).
\label{tk}
\end{equation}

Instead, the spectral density of the conduction electrons has a dip of
the same magnitude.
In STM experiments in the contact regime a peak corresponding to the
spectral density of localized electrons has been measured \cite{Mohr20,Choi16}, as expected for the SSHC on an insulating film. However, one might also have a dip arising from conduction electrons or an asymmetric line shape caused by Fano interference \cite{Ujs00}. In some cases,
even extended atomic orbitals play the dominant role \cite{Joaq21}. However,
in all these cases there is a feature at the Fermi level whose
characteristic energy is easy to identify experimentally and is given by $T_K$.
Assuming the parameters $U$ and $V$ do not vary along the chain, the local variation of $T_K(j)$ from one site to the other would permit to extract the profile of
$\langle \hat{S}_j^z \rangle$ via Eq. \eqref{vn}.

To calculate $V_j$ in the general interacting case, we proceed as follows. In the Kondo problem, it is known that the relevant hybridization processes
are those which  lead to a spin flip between localized spin states $|\uparrow \rangle$ and $|\downarrow \rangle$ (along with an inverted spin flip in the conduction band) in second order \cite{Anderson70}. Then, our general strategy is to extract the effective hybridization in the interacting case from the relation
$\left(V_j/V\right)^2 = \left| \langle \downarrow |  c_{j \downarrow}^\dagger c_{j \uparrow} | \uparrow \rangle \right|$,
obtained from a Schrieffer-Wolff transformation \cite{SW66} neglecting the energy dependence
of the intermediate states.
For a semi-infinite chain, the ground state with spin up can be written as
\begin{align}
|\uparrow \rangle &=a_{0}|1/2\rangle +a_{1\uparrow }c_{j\uparrow }^{\dagger
}|0\rangle +a_{1\downarrow }c_{j\downarrow }^{\dagger }|1\rangle
+a_{2}c_{j\uparrow }^{\dagger }c_{j\downarrow }^{\dagger }|1/2,\beta \rangle
,  \label{up}
\end{align}
where the states $|S_{z},\gamma \rangle $ above are many-body states of the system in the absence of the site $j$, and are classified by their total spin projection $S_{z}$
and, when necessary, by an additional index $\gamma$
to distinguish states with the same $S_{z}$.
Without loss of generality, the coefficients $a_{\nu }$ can be assumed real.
The ground state with spin down is the time reversal of the state above
\begin{align}
|\downarrow \rangle =&a_{0}|-1/2\rangle +a_{1\uparrow }c_{j\downarrow
}^{\dagger }|0,\beta \rangle -a_{1\downarrow }c_{j\uparrow }^{\dagger
}|-1\rangle \nonumber \\
&+a_{2}c_{j\uparrow }^{\dagger }c_{j\downarrow }^{\dagger
}|-1/2,\beta \rangle ,  \label{down}
\end{align}
where each state $|S_{z},\gamma \rangle $ here is the time reversal of the
corresponding one in Eq. (\ref{up}). Note that all of them are orthogonal to
the previous ones because of the different $S_{z}$ except
$|0,\beta \rangle $
and $|0\rangle$. Then
\begin{equation}
(V_{j}/V)^{2}=|\langle \downarrow |c_{j\downarrow }^{\dagger }c_{j\uparrow
}|\uparrow \rangle |=a_{1\uparrow }^{2}|\langle 0,\beta |0\rangle |.
\label{vj2}
\end{equation}

In the case of sufficiently long but finite chain of length $L$, a practical way to calculate $V_j$ is to take into account that by inversion symmetry, the site $j$ can be mapped
onto the site $L+1-j$ and vice versa and that the state with $S_{z}=1$ (-1)
can be constructed from $|\uparrow \rangle $ ($|\downarrow \rangle $) at both
ends. Then, the effective hopping is computed as

\begin{equation}
(V_{j}/V)^{2}=|\langle -1|c_{j\downarrow }^{\dagger }c_{L+1-j\downarrow
}^{\dagger }c_{L+1-j\uparrow }c_{j\uparrow }|1\rangle |^{1/2}.  \label{vj2b}
\end{equation}
Eq. (\ref{vj2b}) is valid for all sites where either the left or right spin end state dominates, but it becomes inaccurate for a few sites near the center of the chain, where both end states have comparable weight.

Finally for a finite chain, there is an effective interaction between the  spins
at both ends which is responsible for the spin gap. In this case, the total effective
Hamiltonian has the form

\begin{equation}
\hat{H}_{\text{eff}}=\hat{H}_{2}+\Delta_{s}\hat{\mathbf{S}}_{L}\cdot \hat{\mathbf{S}}_{R},
\label{heff}
\end{equation}
where $\hat{\mathbf{S}}_{L}$ and $\hat{\mathbf{S}}_{R}$ are the spin operators of both ends.
The essential physics of this model is known from previous studies
of similar systems \cite{Cornaglia05,Dias06,Comment,Reply,Vaugier07,Almeida25, Yuan25}.
If $T_{K} \gg \Delta_{s}$, with $T_{K}$ given by Eq. (\ref{tk}), the spectral
density of the localized states at zero temperature $T=0$ consist
of a peak of half width
$T_{K}$ but with a very narrow dip at the Fermi energy of width
$T_0 \sim T_{K} \exp (-\pi T_{K}/\Delta_{s})$ \cite{Cornaglia05, Yuan25}.
For $T>T_0$ this dip disappears. Instead if $T_{K} \ll \Delta_{s}$,
the spectrum consists of two peaks at energies near $\pm \Delta_{s}/2$.
However, in this regime we expect the ESR-STM technique to describe more
accurately the excitation energy $\Delta_{s}$.
For $T_{K} \sim \Delta_{s}$,
the spectral density consists of two peaks with a dip in between
at the Fermi energy but without a simple quantitative description.

In any case, the effective $T_K$ can be tuned by moving the STM tip vertically with respect to the chain. As the tip approaches the chain, the hybridization $V$, and therefore $\Gamma_j$, increase. Because V changes exponentially with tip-sample distance \cite{Huang20}
, and TK itself depends exponentially on V, it is experimentally straightforward to access the aforementioned limiting regimes, in which either the width of the main spectral feature as a function of position, or the energy of the ESR signal give direct information on the topological state.

\section{Numerical results}
\label{res}

\begin{figure}[t]
 \includegraphics[width=1.0\columnwidth]{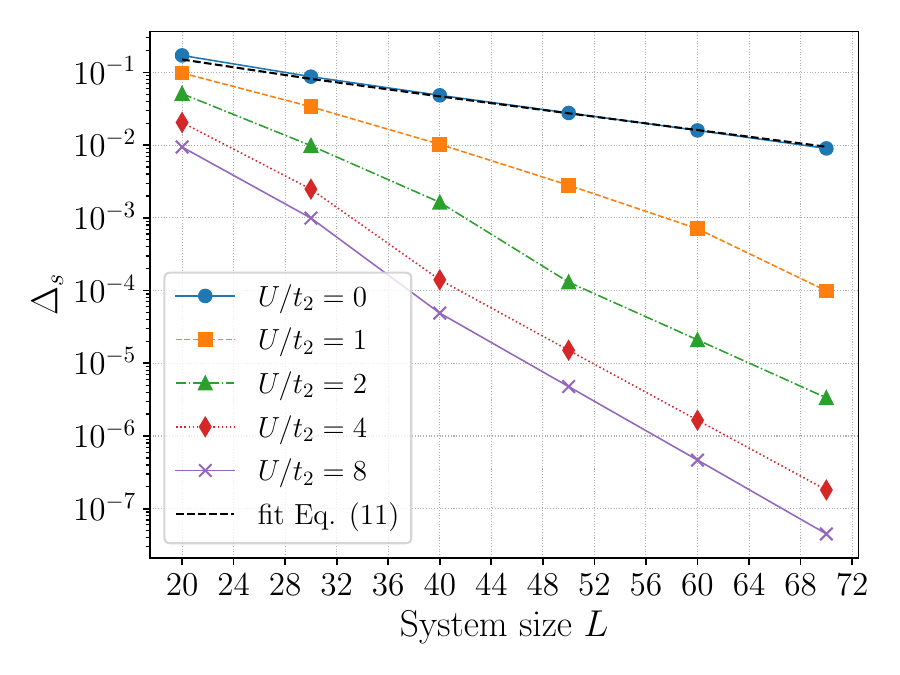}
\caption{(Color online) Spin gap as a function of system size for $t_1=0.9$ and several values of $U$. The dashed line for $U=0$ corresponds to Eq. (\ref{sg}). Dataset available in Ref. \cite{dataset}.}
\label{gaps}
\end{figure}

In this section, we show numerical results for the SSHC (without including the STM tip) which illustrate our proposal. To that end, using the DMRG method implemented through the ITENSOR library~\cite{ITENSOR-PAPER},  we have computed the spin gap $\Delta_{s}$ and the spin profile of the localized spin $\hat{S}_{j}^{z}$ taken in the
lowest-energy state with total spin projection $S_{z}=1$, of interacting play with different lengths $L$ and for different values of $U$.
These are the quantities which permit to predict the STM differential conductance
and the ESR-STM signal according to the previous Section.  All calculations were carried out with a maximum bond dimension sufficient to ensure that the truncation error remained below $10^{-9}$.
This assures that
errors of the DMRG computation are smaller than symbol sizes in each figure.
In what follows we take $t_2=1$ as the unit of energy.

In Fig. \ref{gaps} we show the resulting spin gap, computed as $\Delta_s=E^{(1)}_g-E^{(0)}_g$, where $E^{(S)}_g$ is the ground-state energy in the spin-$S$ subspace, for several lengths of the chain
and different values of the interaction $U$. In the non-interacting case, $U=0$,
the numerical results are in excellent agreement
with the analytical expression given by Eq. (\ref{sg}),
particularly for long chains. There is an
exponential decay with system size. For larger values of $U$ the dependence
on the length of the chain continues to be approximately exponential,
but the decay is faster and the gap is smaller. Increasing $U$ from
0 to 8, the spin gap $\Delta_s$ decreases by slightly more than an order of magnitude
for $L=20$, but by more than five orders of magnitude for $L=70$.
As it was discussed before for spin-1 chains \cite{Mai24} a very small
excitation energy cannot be detected by STM-ESR, unless a Zeeman splitting
is induced by an applied magnetic field. This is precisely the regime (i.e., the case of long chains and large $U$) where our proposed method would be most useful and would complement the standard STM-ESR technique.

\begin{figure}[t]
 \includegraphics[width=1.0\columnwidth]{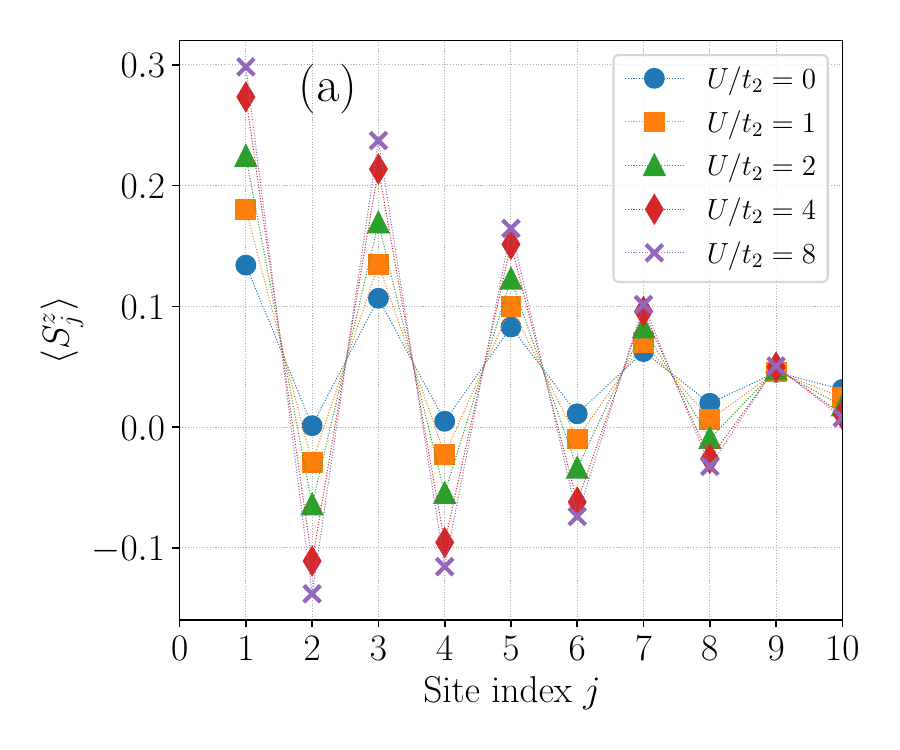}
 \includegraphics[width=1.0\columnwidth]{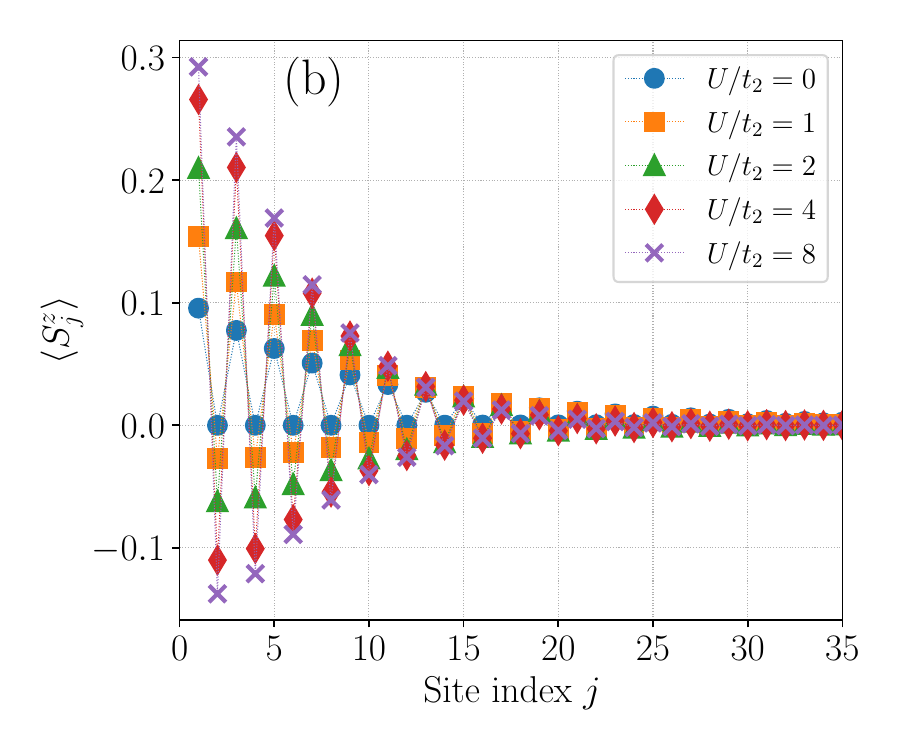}
\caption{(Color online) Expectation value of the spin projection for each site
for a SSHC of 20 sites (panel a) and 70 sites (panel b), $S_z=1$, $t_1=0.9$ and several values of $U$. Due to inversion symmetry with respect with the middle of the chain, only the left half is shown. The complete datasets are available in Ref. \cite{dataset}.}
\label{sz2070}
\end{figure}

In Fig. \ref{sz2070}(a) the expectation values $\langle \hat{S}_{j}^{z}\rangle$ as a function
of site $j$ are shown for the lowest state with total spin projection $S_z=1$, and
a relatively short chain ($L=20$). For $U=0$, the numerical values are in reasonable agreement with the analytical ones obtained from the wave functions
given by Eqs. (\ref{state_l}) and (\ref{state_r}), although for short chains some deviations are expected.
For example for the first site, the analytical
result is $\langle \hat{S}_{1}^{z}\rangle=0.108$, and the corresponding numerical
result is 0.134. As $U$ increases, the edge spin polarization grows, reaching $\langle \hat{S}_{1}^{z}\rangle=0.298$ for $U=8$.
Instead, the neighboring even sites near the left end (and odd sites near the right end) show negligible polarization at $U=0$, but develop finite values for $U>0$ due to the onset of antiferromagnetic
correlations. These correlations are similar to those of the Hubbard model
and arise from effective antiferromagnetic Heisenberg exchange parameters 
$J_1\sim 4t^2_1/U$ and $J_2\sim 4t^2_2/U$. In Fig.  \ref{sz2070}(b) we show similar results for a longer chain ($L=70$). For $U=0$,
$\langle \hat{S}_{j}^{z}\rangle$ at the ends is 0.0956 compared to 0.0951
of the analytical value. In general, the magnitude of the spin projections
decrease slightly with respect to the shorter chain but keeping the same
qualitative behavior. For example
$\langle S_{11}^{z}\rangle / \langle S_{1}^{z}\rangle=0.35$ for $U=0$ and 0.17
for $U=8$. In the middle of the chain ($21<j<50$) all the values verify
$|\langle S_{j}^{z}\rangle|<0.01$.

We note that the faster decay of $\langle \hat{S}_{j}^{z}\rangle$ along the chain in Fig. \ref{sz2070} as $U$ increases is consistent with the results in Fig.~\ref{gaps}, and can be qualitatively understood as a decreasing spin end-state localization length $\xi_s$. While the detailed description of the evolution with $U$ is highly non-trivial, some insight can be obtained in the strongly-interacting limit, where the interacting
SSH model can be mapped  onto the alternating $J_1 - J_2$ Heisenberg model \cite{Hida92_Crossover_Haldane_alternating_Heisenberg_spin_chain, Cross79, Orignac04_Spin_Peierls_Bosonization}. Using the framework of field-theoretic Abelian bosonization \cite{Orignac04_Spin_Peierls_Bosonization}, we can define 
the spinon velocity as 
$v\sim \frac{\pi}{2}\bar{J}a$, where 
$a$ is the lattice parameter and $\bar{J}=(J_1+J_2)/2$,
and using the expression of the singlet-triplet bulk gap 
$\Delta_b\sim\bar{J}\eta^{2/3}/|\ln{\eta}|^{1/2} $ given in those references, where $\eta=|J_1-J_2|/(J_1+J_2) \approx 2 \delta/t_0$ is the dimensionless alternation exchange parameter,
the typical localization length of the topological end states can be estimated as $\xi_s=\hbar v/\Delta_b\sim \pi a \eta^{2/3}/|\ln{\eta}|^{1/2}$, which is independent of $U$. Indeed, in  Fig. \ref{gaps} the curves for $U=4$ and $U=8$ seem to saturate to the same value of the slope, suggesting that the strongly-interacting limit where $\xi_s$ is constant has been reached, in accordance with the previous
argument.

\begin{figure}[t]
\includegraphics[width=1.0\columnwidth]{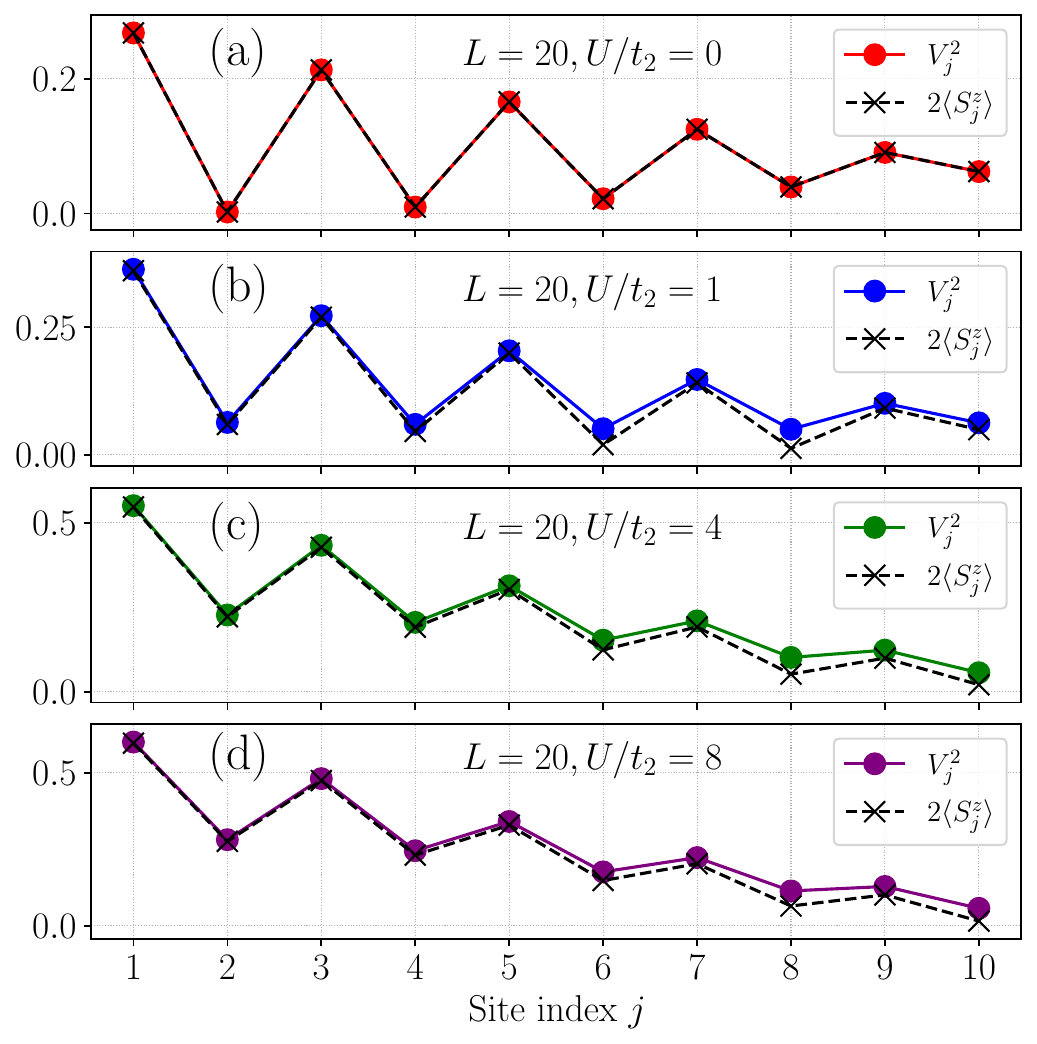}
\includegraphics[width=1.0\columnwidth]{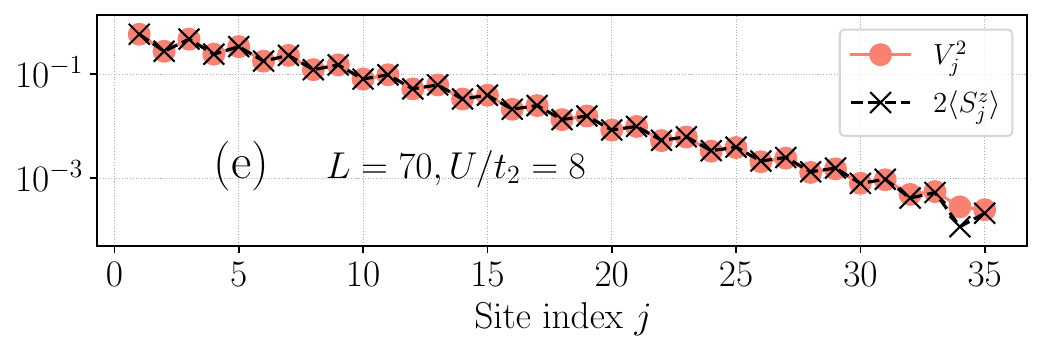}
\caption{(Color online) Comparison between $V_j^2$ calculated using Eq. (\ref{vj2b}) and $2\left|\langle S^z_j\rangle \right|$ obtained from Fig. \ref{sz2070}, for various values of $U/t_2$ ($t_1/t_2=0.9$ is the same for all panels). Panels (a)-(d) correspond to a chain of 20 sites, while panel (e) (note the log scale in the vertical axis) correspond to a 70-site chain and $U/t_2=8$.}
\label{fig:V2_vs_2Sz}
\end{figure}

In Fig. \ref{fig:V2_vs_2Sz}, we compare the effective hybridization determined
numerically with DMRG using Eq. \eqref{vj2b} with our ansatz Eq.
\eqref{vn}, using the values of the spin projections shown in Fig. \ref{sz2070}.
The agreement is remarkable. As stated in the previous Section, Eq. \eqref{vj2b}
is not expected to remain accurate when the spin distributions from the left and right overlap, an effect that occurs near the middle of the chain. This likely explains the small deviations observed close to the center of the 20-site chain. For the 70-site chain, this overlap is reduced, and the agreement between the two results holds throughout most of the system, except very near the center, even though the effective hybridization spans more than three orders of magnitude.

\begin{figure}[h!]
\includegraphics[width=0.95\columnwidth]{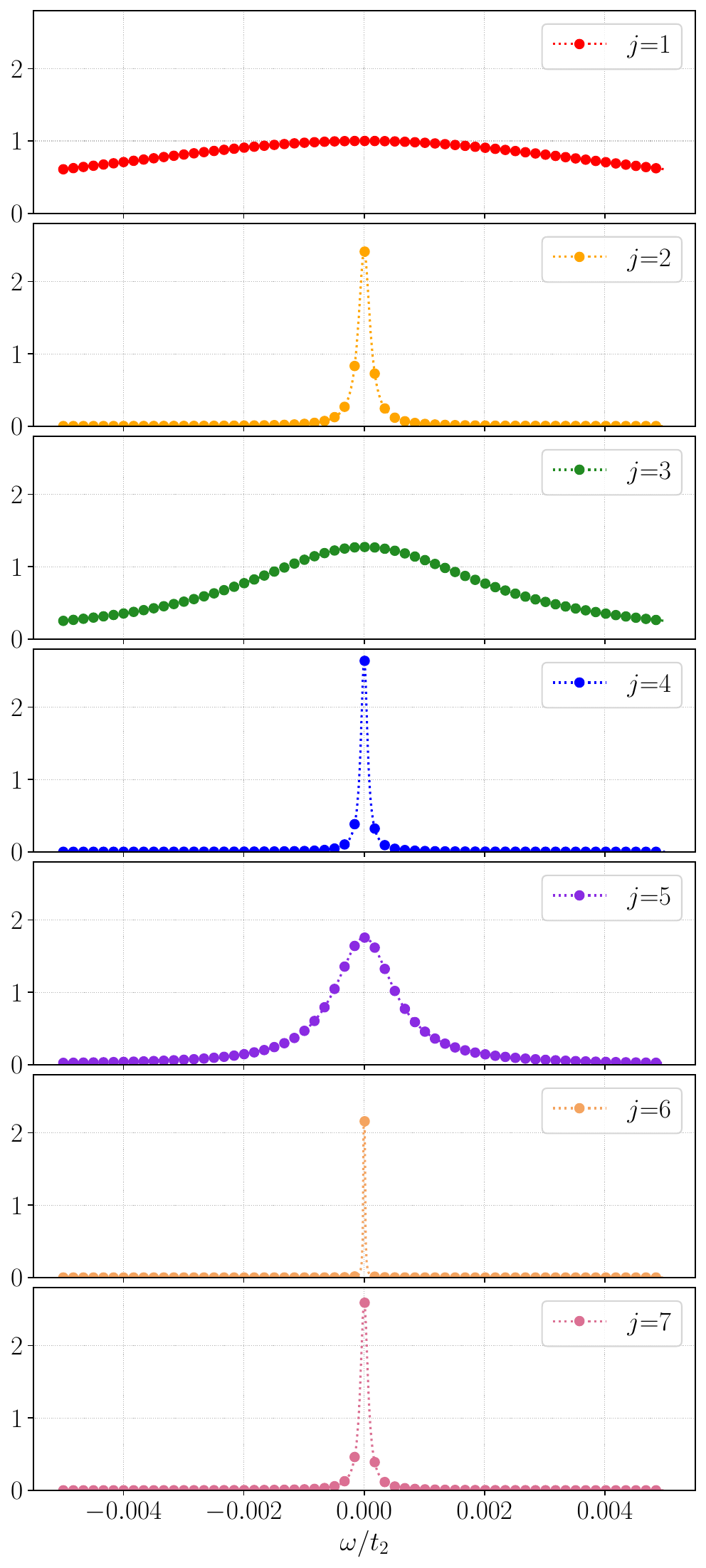}
\caption{(Color online) LDOS at the impurity site $j$ normalized to the first site of the chain at $\omega=0$, $\rho_j(\omega)/\rho_0(0)$, vs $\omega$ obtained from Eq. (\ref{ldos})  for a chain of $L=20$ sites. The values of $ V^2_j$ used correspond to those used in  Fig. \ref{fig:V2_vs_2Sz}(c). Other parameters are $U=4 t_2$ and $\Gamma_0=20 t_2$ (chosen for illustrative purposes).}
\label{rhoj}
\end{figure}

Having computed the spin gap and spin profile of the isolated SSHC, and having shown that the latter controls the effective hybridization, we now describe how these quantities can be extracted from STM experiments.
The differential conductance at the position of the $j^{\text{th}}$ impurity,
$\mathbf{r}_j$, for a chain
sufficiently isolated from the metallic substrace is expected to be dominated by
the localized spectral density \cite{Mohr20,Choi16}, i.e. $dI/dV(\mathbf{r}_j , \omega)\sim \rho_j(\omega)$, where
\begin{align}
    \rho_j(\omega)&=-\frac{1}{\pi} \textrm{Im} \sum_{\sigma} \langle
\langle c_{j\sigma };c_{j\sigma }^{\dagger }\rangle \rangle _{\omega },
\label{ldos}
\end{align}
is the local density of states (LDOS) in the SSHC at site $j$, and where $\langle
\langle c_{j\sigma };c_{j\sigma }^{\dagger }\rangle \rangle _{\omega }$ is the corresponding retarded Green's function at that site. In the Kondo regime, when the conduction electrons in the STM tip effectively screen the $j^{\text{th}}$ spin in the SSHC, the retarded Green's function can be reasonably well approximated by the expression \cite{hewson, Agam01}

\begin{align}
    \langle
\langle c_{j\sigma };c_{j\sigma }^{\dagger }\rangle \rangle _{\omega } \sim \frac{Z_{K,j}}{\omega-E_F+iT_K(j)},
\end{align}
where $Z_{K,j}=T_K(j)/\pi \Gamma_j$ is the quasiparticle weight, $T_K(j)$ is the position-dependent Kondo temperature given in Eq. (\ref{tk}), and $E_F$ is the Fermi energy (taken as the zero of energies). In Fig. \ref{rhoj} we illustrate the predicted $\rho_j(\omega)$ (normalized by the LDOS at the first site at the Fermi energy), using as input the spin profile obtained in Fig. \ref{sz2070}(a). Due to inversion symmetry, only the left half of the SSHC is shown. For illustrative purposes, we have assumed parameters $U/t_2=4$ and $\rho V^2/t_2=20$. We note that the width of the curves vary considerably from site to site in the chain, indicating a strong variation of
$T_K(j)$.
As mentioned in the previous Section, the line shape can be altered in presence of interference effects with other orbitals \cite{Ujs00,Joaq21}. However in all
cases, the characteristic energy scale $T_K(j)$ can be extracted.

\begin{figure}[t]
\includegraphics[width=1.0\columnwidth]{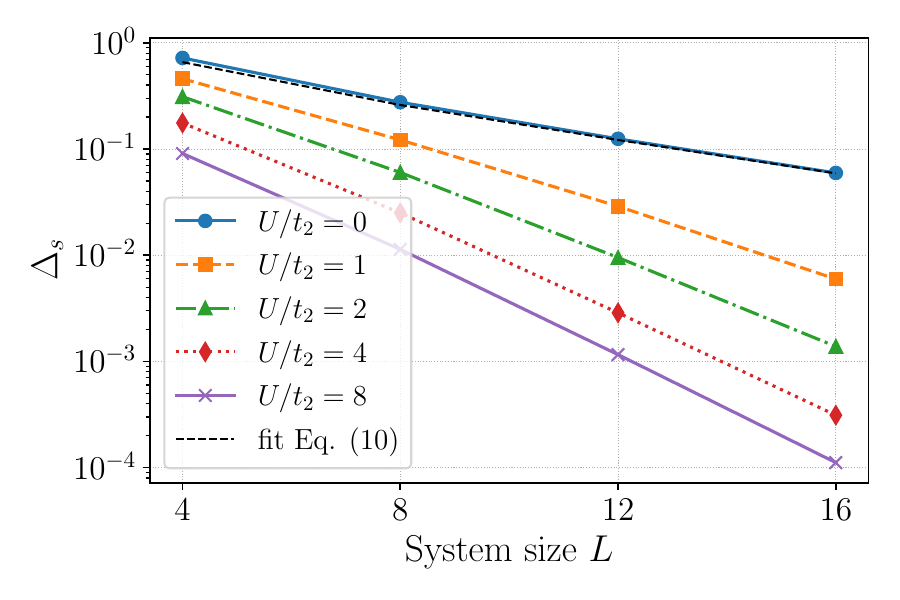}
\caption{(Color online) Spin gap as a function of system size for $t_1=0.7$ and several values of $U$. The dashed line for $U=0$ corresponds to Eq. (\ref{sg}). Dataset available in Ref. \cite{dataset}.}
\label{gapsp7}
\end{figure}

In the rest of this Section we discuss how the results presented above
change with the difference between hoppings. From the form of the wave
functions in the non interacting case, Eqs. (\ref{state_l}) and (\ref{state_r}), one can define a decay length $\lambda= 1/\text{ln}(t_2/t_1)$, which decreases with
decreasing $t_1$ and diverges at the limit of the topological phase
$t_1 \rightarrow t_2$.

In Fig. \ref{gapsp7} we show the spin gap $\Delta_s$ for a smaller value of
$t_1$ (0.7) as that used in Fig. \ref{gaps} already shown.
The same values of the interaction $U$ are used in both figures,
but smaller lengths in Fig. \ref{gapsp7}, according to the expected
reduction in $\lambda$. Qualitatively, the behavior of $\Delta_s$
is similar in both figures.
Again, in the non-interacting case, $U=0$,
the numerical results are in very good agreement
with the analytical expression Eq. (\ref{sg}), but small
deviations are present for the smallest sizes.

\begin{figure}[t]
\includegraphics[width=1.0\columnwidth]{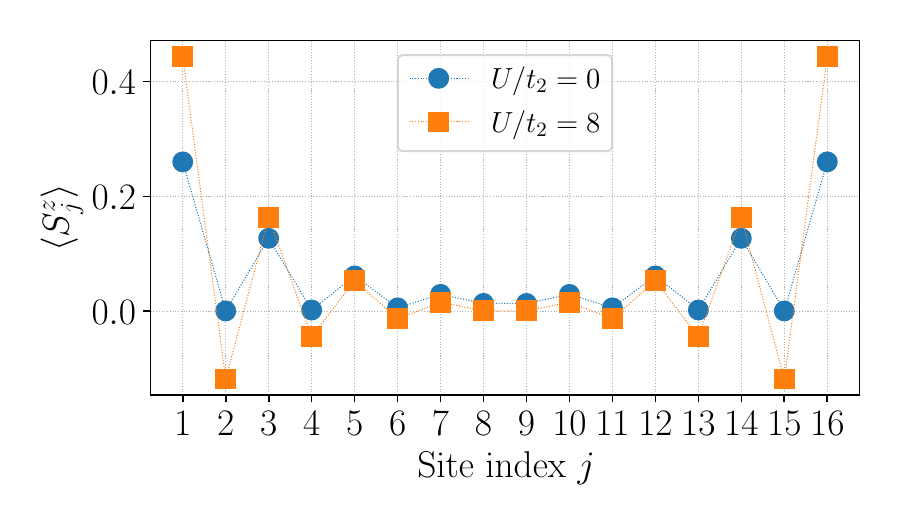}
\caption{(Color online) Expectation value of the spin projection for each site
for a SSHC of 16 sites, $S_z=1$, $t_1=0.7$ and two values of $U$. Dataset available in Ref. \cite{dataset}.}
\label{sz16}
\end{figure}

In Fig. \ref{sz16}, we show the expectation values $\langle \hat{S}_{j}^{z}\rangle$ as a function of the site-index $j$  for the lowest state with total spin projection $S_z=1$. The result is qualitatively similar to those of  Figs. \ref{sz2070} (a) and (b). For $U=0$, the analytical
result is $\langle S_{1}^{z}\rangle=0.256$, and the corresponding numerical
result is 0.266.

\section{Zero-frequency Green’s function}\label{green}

The topological classification of interacting gapped phases often relies on invariants constructed from the single-particle Green’s functions. These invariants capture, on equal footing, the appearance of zero-energy boundary poles—associated with genuine single-particle excitations—and of edge-localized zeros of the determinant of the Green’s functions, which reflect poles of the self-energy. Several works have suggested that such zeros may correspond to neutral excitations lacking the full set of electron quantum numbers~\cite{Wagner23,Fabrizio23,Pangburn25}.

These zeros are also closely related to the topological aspects of the Luttinger theorem.
Seki and Yunoki~\cite{Seki17} have shown that the difference in the Luttinger volume of interacting and non-interacting systems is precisely given by the Luttinger integral $I_L$, which can take a non-trivial value when the determinant of the Green’s function exhibits a zero at $\omega=0$, in contrast to the conventional Luttinger theorem, which states that $I_L=0$
under the assumption that perturbation theory in the interaction is valid to all orders \cite{Lut60,LW60}. Logan and Galpin find a jump in $I_L$ at the metal-insulator transition in the Hubard model \cite{Logan16}.
%A jump in $I_L$ might naturally explain the observed change in the
%volume enclosed by the Fermi surface in the cuprates from a large
%one (corresponding to the non-interacting system) to a small one,
%with one particle less \cite{Chan25}.
Moreover, non-trivial, topological values of the Luttinger integral have been shown to play an important role in impurity systems \cite{Curtin18,Blesio18,Zitko21,Blesio23,BlesioB,Blesio24}.
Therefore, the study of the zeros of the determinant of the Green’s functions at zero energy is of fundamental interest.

In the SSH–Hubbard model considered here, it is thus natural to ask whether boundary zeros of the propagator in the topological phase can be identified with the localized free spins, which are susceptible to Kondo screening once coupled to a metallic bath. A closely related argument has been made for the half-filled Hubbard atom~\cite{Blason2023}, which corresponds to the trivial edge limit of the fully dimerized ($t_1=0$) SSH–Hubbard chain. As we will discuss in this Section, zero eigenvalues of the propagator emerge in the thermodynamic limit of the topological phase; however, their corresponding eigenvectors are edge-localized with a characteristic length scale that differs substantially from that of the boundary spin degrees of freedom.

The properties of the Green's functions of the isolated SSH-Hubbard chain at finite $U$ and near zero frequency can be analyzed using a combination of analytical
and numerical tools. The K\"all\'en–Lehmann spectral representation of the Matsubara propagator in the zero temperature limit takes the form

\begin{eqnarray}
G_{ij}(i\omega) &=&\sum\limits_{\nu }\frac{D_{ij}^{\nu }}{i\omega  +E_{\nu }-E_{g}}  \notag \\
&&+\sum\limits_{\mu }\frac{C_{ij}^{\mu }}{i\omega  -E_{\mu }+E_{g}},
\label{gij}
\end{eqnarray}
where

\begin{eqnarray}
D_{ij}^{\nu } &=&\langle g|c_{j\sigma }^{\dagger }|\nu \rangle \langle \nu
|c_{i\sigma }|g\rangle ,  \notag \\
C_{ij}^{\mu } &=&\langle g|c_{i\sigma }|\mu \rangle \langle \mu |c_{j\sigma
}^{\dagger }|g\rangle   \label{cd}.
\end{eqnarray}
Here, $|\nu \rangle ,E_{\nu }$ ($|\mu
\rangle ,E_{\mu }$) are eigenstates and the corresponding energies for one
particle less (more) than the half-filled case and $|g\rangle $ is the
ground state assumed non-degenerate. For a Kramer's degenerate ground state $|g\rangle $ is taken as a mixture of equal weight of both states.

As a consequence of the electron-hole symmetry $S$ in the model at half
filling (specifically the substitution $c_{j\sigma }^{\dagger }\rightarrow
(-1)^{j}c_{j\sigma }$ leaves the model invariant,  $SHS=H$,   $Sc_{j\sigma
}^{\dagger }S=(-1)^{j}c_{j\sigma }$), we can choose in Eq. (\ref{gij}) $|\mu
\rangle =S|\nu \rangle $, implying $S|\mu \rangle =|\nu \rangle $) and
$E_{\mu }=E_{\nu }$. Also $S|g\rangle =|g\rangle $.
Then from Eq. (\ref{cd})

\begin{equation}
C_{ij}^{\mu }=-(-1)^{j-i} \langle g|c_{i\sigma }^{\dagger }|\nu \rangle \langle \nu
|c_{j\sigma }|g\rangle =-(-1)^{j-i}D_{ji}^{\nu }.  \label{cd2}
\end{equation}
We note that because of the presence of a charge gap at finite $U$, there are no poles at zero frequency, and therefore one can safely evaluate $\omega =0$ in Eq.~\eqref{gij}. Using Eq.~\eqref{cd2}, we obtain

\begin{equation}
G_{ij}(0)=\sum\limits_{\nu }\frac{D_{ij}^{\nu }-(-1)^{j-i}D_{ji}^{\nu }}{E_{\nu
}-E_{g}}.  \label{gij2}
\end{equation}
Clearly, from Eq. (\ref{gij2}),

\begin{equation}
G_{ji}(0)=-(-1)^{j-i}G_{ij}(0).  
\label{gji}
\end{equation}

For a system with time-reversal symmetry and without spin-orbit coupling one
expects the Lehmann amplitudes in Eq.~\eqref{cd} to be real. Together with Eq.~\eqref{gji}, this implies that the zero-frequency Green’s function matrix $G(0)$, with elements $G_{ij}(0)$, is real, symmetric, and vanishes for even $j-i$, consistent with chiral sublattice symmetry. A more stringent result was recently reported in Ref.~\cite{Lehmann2025} for the large-$U$ limit, where $G(0)$ acquires a tridiagonal structure, coupling only nearest-neighbor sites. Quite generally, the resulting structure of $G(0)$ is that of an effective noninteracting tight-binding model with electron–hole symmetry $S$. Consequently, each positive eigenvalue has a corresponding negative partner. For a chain with an odd number of sites, this symmetry enforces the presence of a zero eigenvalue, independently of the ratio $t_1/t_2$. For a chain with an even number of sites, two eigenvalues approach zero in the thermodynamic limit of the topological phase ($t_1<t_2$). This is consistent with the numerical evaluation of the bulk topological invariant reported by Manmana {\it et al.} \cite{Man12}.

To verify the predictions of the symmetry analysis above, we performed a numerical check of the Green's function matrix for a small chain of seven sites with $U=1$, $t_{2}=1$, and several small $t_{1}$ values. For each of the two degenerate $S_{z}=\pm 1/2$ ground states we constructed the corresponding propagators and then averaged them. In the spin-polarized cases, sizable matrix elements connect sites within the same sublattice, but these vanish upon averaging, in agreement with the symmetry considerations above. For small $t_1$, we find that the off-diagonal element $G_{12}(0)$ scales as $\sim t_{1}^{3}$, consistent with perturbative arguments. Consequently, the zero-eigenvector of the propagator---the ``dark orbital"---is strongly localized at the first site, in sharp contrast with the boundary spin distribution, which, as shown in Figs.~\ref{sz2070} and \ref{sz16}, extends with appreciable weight to the third site and scales as $\sim t_1^2$. We further note that breaking electron–hole symmetry, for instance by shifting the on-site energy, displaces the zeros of the Green's function away from $\omega=0$ (see Supplemental Material of Ref.~\cite{Wagner23}), while leaving the localized end spins unaffected.

It is interesting to analyze the contribution of the states near the Fermi
energy for the non-interacting case. Using the results of Section \ref{zm}
we obtain for odd $j$
\begin{equation}
G_{jj}(\omega)=\frac{z^{j-1}(1-z^{2})}{2(1-z^{L})}\left( \frac{1}{\omega +t_{LR}} +\frac{1}{\omega -t_{LR}}\right) .  \label{gni}
\end{equation}
For an infinite chain $t_{LR}\rightarrow 0$, and the Green functions have
a pole at $\omega =0$. For a finite chain, the pole is converted into a
zero. However, this zero does not imply a vanishing of the determinant of
one-particle Green functions, which is the quantity
related with topology \cite{Gura11,Man12,Wagner23}.

In addition, because of the wavefunction decay factor $z^{j-1}$ [see Eq. (\ref{op_lsigma})], the low-energy features are
substantially more marked at the end of the chain, where also larger
expectation values of the spin projection are present.

\section{Summary and discussion}

\label{sum} 

In this work we provide a theoretical framework
for probing the spins localized near the ends of an
interacting Su–Schrieffer–Heeger chain
in the topological phase, using scanning tunneling microscopy in the contact regime, combined with electron spin resonance in the tunneling regime.

We propose that when the STM tip couples directly to these localized spins, the system maps onto an effective Anderson impurity model, producing a Kondo resonance whose spatially dependent width encodes the spin profile of the end states.

While for
a non-interacting infinite open chain vanishingly small charge and spin excitation gaps occur in the lowest-energy subspace due to the presence of zero-energy states localized at the ends, in the interacting case a finite charge gap opens (albeit with a vanishingly small spin gap) in the infinite open chain.
In other words the electronic zero-energy excitations (with charge and spin)
in the non-interacting case $U=0$ disappear for finite $U$ despite the fact that
zero-energy spin excitations remain in the infinite chain. For finite chains
the end states interact leading to a finite spin gap which, as we have described,
can be measured combining STM and ESR.

Using density-matrix renormalization group, we compute the spin gap and spin polarization for chains of different lengths and interaction strengths, showing how correlations reduce the spin gap and enhance spin localization at the edges. For $U=0$ we obtain analytical results for the spin gap and distribution of the
spin projection $\langle S_{j}^{z}\rangle$ along an open topological SSHC which agree well with DMRG results for sufficiently long chains. In the interacting
case, the magnitude of the spin projection and its dependence on position
become evident in the numerical calculations choosing the subspace with total spin projection $S_z=1$.
The numerical results for these two quantities and its dependence on parameters clarify the nature of the spins at the end states and might serve as a guide for future research.

As a consistency check with previous works defining topology via single-particle Green's function invariants, we verified that the zero-frequency propagator respects chiral sublattice symmetry, which enforces the presence of zero eigenvalues in the topological regime. We numerically analyzed the corresponding eigenvectors, finding them strongly localized at the chain ends, in contrast with the more extended distribution of the boundary spins.

Taken together, our results establish STM and ESR as powerful probes of correlated topological boundary excitations, providing a promising route to characterize boundary spins in interacting SSH chains.

Our proposal might be useful to detect magnetic impurities in Heisenberg
chains \cite{Eggert92}.
It would also be interesting to extend our theory to systems which have fractional
spin end modes, like the XXZ spin-1/2 model \cite{Kattel1,Kattel2} and
time-reversal invariant topological superconductors
\cite{Keselman13,Aligia18,Aligia19}.

\section*{Acknowledgments}

We thank Pablo Cornaglia for useful discussions. L.P.G. acknowledges support provided by the FRS-FNRS Belgium and the L'Or\'eal-UNESCO for Women in Science Programme. The results presented in this work have been obtained by using the facilities of the CCT-Rosario Computational Center, member of the High Performance Computing National System (SNCAD, MincyT-Argentina).

\bibliography{ref2}

\end{document}